\documentclass[prd,aps]{revtex4}
\usepackage{graphicx}
\usepackage{epsfig}
\fussy
\def\be{\begin{equation}}
\def\ee{\end{equation}}
\def\bea{\begin{eqnarray}}
\def\eea{\end{eqnarray}}
\def\barr{\begin{array}}
\def\earr{\end{array}}
\def\dis{\displaystyle}

\def\square{\hbox{\vrule width 0.4pt height0.5em depth0.1em\kern-0.4pt
\raise0.5em\hbox{\vrule width 0.6em height.32pt depth 0.08pt}\kern
-0.4pt\vrule width 0.4pt height0.5em depth0.1em\kern-0.6em\lower0.1em\hbox
{\vrule width 0.6em height.32pt depth 0.08pt}}}
\def\Box{\relax\ifmmode\square\else\nobreak\null\hfill$\square$\fi}
\let\Box

\newcommand{\newc}{\newcommand}
\newc{\gsim}{\lower.7ex\hbox{$\;\stackrel{\textstyle>}{\sim}\;$}}
\newc{\lsim}{\lower.7ex\hbox{$\;\stackrel{\textstyle<}{\sim}\;$}}
\newc{\gev}{\,{\rm GeV}}
\newc{\mev}{\,{\rm MeV}}
\newc{\ev}{\,{\rm eV}}
\newc{\kev}{\,{\rm keV}}
\newc{\tev}{\,{\rm TeV}}

\newc{\ie}{{\it i.e.}}
\newc{\etal}{{\it et al.}}
\newc{\eg}{{\it e.g.}}
\newc{\etc}{{\it etc.}}
\newc{\cf}{{\it c.f.}}
\begin{document}

\catcode`@=11 \@addtoreset{equation}{section} \catcode `@=12
\renewcommand{\theequation}{\thesection.\arabic{equation}}

{}~ \hfill\vbox{\hbox{hep-th/0103248}\hbox{MRI-P-010304}
}%

\vskip 2mm

\title{
Brane Dynamics in the Randall-Sundrum model, %
Inflation and Graceful Exit}

\vspace*{1ex}

\author{\large \rm
Somdatta Bhattacharya$^{\ddagger}$,
Debajyoti Choudhury$^{\star}$,
Dileep P. Jatkar$^{\star}$ and
Anjan Ananda Sen$^{\star}$
\footnote{E-mail:  som@tnp.saha.ernet.in, debchou@mri.ernet.in,
dileep@mri.ernet.in, anjan@mri.ernet.in
}}

\vspace*{2ex}

\affiliation{{$^{\ddagger}$ 
Theoretical Physics Division, 
Saha Institute of Nuclear Physics,
1/AF, Bidhan Nagar, Kolkata 700 064, INDIA}\\
$^{\star}$
Harish-Chandra Research Institute
\footnote{Formerly, Mehta Research Institute of Mathematics
and Mathematical Physics.}, 
Chhatnag Road, Jhusi, Allahabad 211 019, INDIA}
\vspace*{1ex}

\begin{abstract}
  We study  the averaged action  of  the Randall-Sundrum model with  a
  time dependent  metric ansatz. It can  be reformulated in terms of a
  Brans-Dicke action with   time dependent Newton's constant.  We show
  that the physics of early    universe, particularly inflation,    is
  governed by the Brans-Dicke theory. The Brans-Dicke scalar, however,
  quickly  settles to  its  equilibrium value  and decouples from  the
  post-inflationary cosmology.  The deceleration parameter is negative
  to start with but changes sign before the Brans-Dicke scalar settles
  to  its equilibrium value.  Consequently,  the brane metric smoothly
  exits inflation.   We have also studied  the slow-roll  inflation in
  our  model and investigated  the spectra of the density perturbation
  generated by  the radion field  and find  them  consistent with  the
  current observations.
\end{abstract}

\maketitle

\section{Introduction}

The quest  to explore physics beyond  the Standard  Model (SM) has, in
the past couple of years, been dominated by the  idea that space time
is of  a  dimension  larger than four   and  that we  are  essentially
confined to  a  4-dimensional  hypersurface thereof.     Although such
suggestions go back to the work  of Akama~\cite{Akama} and Rubakov and
Shaposhnikov\cite{Rubakov:1983bb,Rubakov:1983bz}, the present interest
has  been   occasioned    by  the  recognition   that   such solitonic
hypersurfaces (or branes) abound in string theory.  While an exact and
viable low-energy realization of string theory is  still some way off,
considerable excitement has been aroused by the prospect of explaining
the embarassing gap between the electroweak scale  $M_{\rm W}$ and the
Planck scale  ($M_{\rm   pl}$, $M_{\rm pl}/M_{\rm  W}\sim10^{16}$)  by
invoking           the       presence     of             such    extra
dimensions\cite{Arkani-Hamed:1998rs,Antoniadis:1998ig,Arkani-Hamed:1999nn,
Randall:1999ee,Randall:1999vf}.  What is  particularly  new of  such
models is that the ordinary matter is confined to our four-dimensional
brane while the gravity propagates in the whole spacetime.

These models can be broadly divided into two sets.
The first, stimulated primarily by the papers of Arkani-Hamed \etal
\cite{Arkani-Hamed:1998rs,Antoniadis:1998ig,Arkani-Hamed:1999nn}
(ADD)(see also \cite{Antoniadis:1990ew}), invokes large (possibly sub-mm)
new spatial dimensions transverse to the SM brane, with the
higher-dimensional metric being factorizable and essentially flat.
With such a construction in place, it can be shown that the
fundamental gravitational scale could be close to the TeV scale,
with the aforementioned hierarchy being explained by the large volume of
the compactified dimensions.

Subsequently, Randall and Sundrum~\cite{Randall:1999ee,Randall:1999vf}
introduced an important variation, wherein the directions parallel and
transverse to the branes do not factorize into a product space. An
immediate consequence is that the 4-dimensional graviton wavefunction
has a non-trivial dependence on the transverse coordinate(s).
If this dependence is of the `correct' form, the bulk (\ie\
higher dimensional)
graviton would be localized away from the brane that we live on, resulting
in a suppression of gravitational interactions in our world.
While such `warped' compactifications had already been
considered in the
literature\cite{Rubakov:1983bz,vanNieuwenhuizen:1985ri}
it was only after RS incorporated the modern brane
idea, that the activity started in right earnest.

The fact that the RS models have far less striking phenomenological
consequences (especially in collider experiments or in astrophysical
contexts) as compared to the those for ADD models could have been
expected to lead a gradual dissolution of interest in such models.
That this has not been the case is, to a great extent, due to
the fact that RS-models are perhaps more easily embeddable in
supergravity and superstring compactifications\cite{HW,CS}.
A particular example is afforded by the heterotic M theory,
whose field theory limit is the 11-dimensional supergravity
compactified on $S_1/Z_2$ with Yang-Mills, albeit supersymmetric,
fields living on the two boundaries\cite{HW}.
Furthermore, on account of their warped geometry, the RS models are able
to resolve the SM hierarchy problem without needing to introduce large
dimensions (a hierarchy problem in itself).
At a more prosaic level, it has been argued that
the early universe evolution of the RS
models is dramatically different from that of standard FRW cosmology. It
is this aspect that we intend to focus on in this artcle.

As is well known, Big Bang cosmology, while successful in answering
many questions, leaves many others unanswered. Prominent amongst these
are the flatness problem, the observed low density of monoples and the
horizon problem.
Theories with inflation (a period in the distant past characterized by an
exponential growth of the scale factor) offer solutions to all of these,
and in fact are the only ones to do so. Inflation, however, is associated
with its own set of problems. Primary amongst these is the generation of
a correct (scalar) potential that would drive inflation, and equally
importantly, the mechanism of a noncontrived exit of the universe
from an inflationary phase. This {\em graceful
exit} problem has been plaguing both cosmologists and string theorists
with no simple solution in sight. Naturally, questions such as this reside
at the core of our investigations.

The cosmological implications of models with extra dimensions have
been discussed by many authors \cite{BDL,CGKT,CGS}.  Inflationary
solutions were obtained for both flat bulk geometry \cite{LOW}, as
well as for a AdS bulk geometry \cite{Nihei,Kaloper,Inflation,more,CGRT}.
Inflation, in
our model, occurs though in a novel way. To be specific, we use the
dynamics of the warp factor to drive inflation.  Starting from the
effective four-dimensional action, recast in terms of the variables of
the negative tension brane, we find that, after some field
redefinitions, it could be interpreted as an action for scalar-tensor
gravity.  Solving the consequent equations of motion, we find that the
evolution is governed by an initial exponential inflationary phase
followed by a decelerating phase at the end of which the radion
stabilises at its present day value.  In other words, the exit is
indeed graceful. Since the warp factor itself is the `inflaton' in
this theory, neither the adequate growth of the scale factor nor the
exit demands any unnatural tuning of parameters.

In section II, we will show that the  averaged four dimensional action
obtained by integrating out the fifth dimension $y$\cite{CGRT}, can be
written as  an action for a  generalised Brans-Dicke theory \cite{BD}. 
The function $\omega(\phi)$ in this theory turns out to be
\be
\omega(\phi) = -\frac{3}{2}\frac{\alpha\phi}{1+\alpha\phi}.
\ee
In section III, we obtain a solution to the Brans-Dicke equations of
motion.  Solution to the equations of motion is such that the
Brans-Dicke (BD) scalar rapidly approaches its equilibrium value. The
universe on the visible brane, which is also known as the standard
model brane, undergoes exponential inflation during this time. As the
scalar settles down to its equilibrium value, its equations of motion
decouples and we effectively get four dimensional Einstein
equations. We also study the behaviour of the deceleration
parameter. We find that it is negative
at early times but that its sign changes well before the
BD scalar stabilizes. This, in turn, implies that the
universe on the visible brane exits the inflationary era.

\section{Brans-Dicke Gravity on the Brane}

The Randall-Sundrum model that we will be interested in this paper consists of
five dimensional Einstein gravity with negative cosmological constant and two
3-branes located at the fixed points of the orbifold $S^{1}/Z_{2}$. We
parametrize the orbifold direction $y$ in such a way that the orbifold fixed
points are at $y = 0$ and at $y = 1/2$ where the
positive tension and negative tension branes are respectively located.
The action\footnote{Our metric signature is $(-,+,+,+,+)$.} is given by
\cite{Randall:1999ee,Randall:1999vf}
\be
S = 2 \int d^{4}x
\int^{1/2}_{0}dy\sqrt{-G}(M_{5}^{3}R-\Lambda+{\cal{L}}_{R}) + \int d^{4}x
\sqrt{-g^{(+)}}(L^{+}-V^{+}) + \int d^{4}x\sqrt{-g^{(-)}}(L^{-}-V^{-})
\label{5d-action}
\ee
where $G_{MN}\, (M,N=\mu,y)$ is the five dimensional bulk metric, $R$ is the
five dimensional Ricci scalar and
$\Lambda$ is the cosmological constant in bulk,
$M_{5}$ is five dimensional Planck mass. ${\cal{L}}_{R}$ is some non-specified
bulk dynamics responsible for generating a potential for the radion.
The four dimensional metrics $g^{(+)}_{\mu\nu}$ and $g^{(-)}_{\mu\nu}$,
relevant to the  positive
and negative tension brane respectively, can be expressed in terms
of $G_{MN}$ as
\be
g^{(+)}_{\mu\nu} = G_{\mu\nu}(x^{\mu},y=0),\qquad
g^{(-)}_{\mu\nu} = G_{\mu\nu}(x^{\mu},y=1/2).
\label{pmmetric}
\ee
This restriction can be implemented through delta-functions
which fix the locations of two 3-branes.
While $L^{+}$ and $L^{-}$ are the
Lagrangians for the matter fields confined to the
positive and negative tension branes respectively, $V^{\pm}$
correspond to the associated brane tensions.

Equations of motion, in the absence of matter on either brane, are solved
for the metric\cite{Randall:1999ee,Randall:1999vf}
\be
ds^2 = e^{-2m_0r_c|y|}(\eta_{\mu\nu}dx^{\mu}dx^{\nu})+ r_c^2dy^2,
\label{RS-met}
\ee
provided the brane tensions and the bulk cosmological constant are
related through
\be
V^{+} = -V^{-} = 12m_0M_{5}^3,\qquad
\Lambda = -12m_0^2M_{5}^3 \ .
\label{tensions}
\ee
This metric, as is well known by now, gives rise to an exponential hierarchy
between the natural mass scale on the positive tension brane and that on the
negative tension brane. For an appropriate choice of $m_0r_c$, it is
possible to generate the Planck mass $M_{pl}$ on the negative tension
brane to be of the order of $10^{19}$ GeV starting from a bulk energy
scale of about a few TeV.  What is of prime interest is that the
resolution of the hierarchy (between the four dimensional Planck mass
and the electroweak scale) does not necessitate choosing unnaturally
small or big values for any of the parameters in the theory.

We would like to see what kinds of cosmological models we can accomodate in
this scenario. Since our purpose is to study cosmological evolution, we will
consider a new ansatz for the metric where the metric components are functions
of the coordinate $y$ as well as functions of time $t$. We will consider
only a minimal modification of the RS static brane ansatz to accomodate
temporal dependence\cite{CGRT}. In other words, we now have
\be
ds^{2} = e^{-2m_0b(t)|y|}\left[g_{\mu\nu}dx^{\mu}dx^{\nu}\right] +
b^{2}(t)dy^{2}
\label{RS-time}
\ee
where 
\be
g_{\mu\nu} = {\rm diag}\left(-1, a^{2}(t), a^{2}(t), a^{2}(t)\right).
\label{hubble}
\ee
Clearly, this choice is such that when $b(t)\rightarrow {\rm const.} = b_{0}$
and 
$a(t) \rightarrow {\rm const.} =1$, we recover the static RS solution.

The metrics on the two branes can be readily obtained from this ansatz:
\be
g^{(+)}_{\mu\nu} = g_{\mu\nu},\quad
g^{(-)}_{\mu\nu} = e^{-m_{0}b(t)}g_{\mu\nu}.
\label{conformal}
\ee 
It is worth pointing out that the metric on the negative tension brane is
related to that on the positive tension brane by a conformal transformation.
This has an interesting consequence for brane inflation, which we will 
discuss in the next section. We can now use this ansatz to determine
equations of motion for $a(t)$ and $b(t)$. To get these equations let us first
substitute (\ref{RS-time}) and (\ref{hubble}) into (\ref{5d-action}). After
performing the $y$ integration we get the four dimensional effective action 
\cite{CGRT}
\be
S_{eff} = {\frac{3}{\kappa^{2}m_{0}}}
\int d^{4}x \: a^{3}\left[(1-\Omega_{b}^{2})\frac{{\dot{a}}^{2}}{a^{2}} + m_{0}
\Omega_{b}^{2}\frac{\dot{a}}{a}\dot{b} - \frac{m_{0}^{2}}{4}\Omega_{b}^{2}
{\dot{b}}^{2}-V_{r}(b)\right] + \int d^{4}x \: a^{3}L^{-}\Omega_{b}^{4},
\label{4d-action}
\ee
where $\kappa^2 = 1/2M_{5}^3$, $\Omega_{b} \equiv e^{-m_{0}b(t)/2}$ and
$a^{3}V_{r}(b) = -b(t)\int dy\Omega^4_{b}{\cal{L}}_{R}$. We have ignored 
the matter term in the positive tension brane but have considered 
one on the negative tension where our observable universe resides. 
A straightforward calculation shows that the action (\ref{4d-action}) can now 
be written as
\be
S_{eff} = \frac{1}{2\kappa^2m_0}
\int d^4x \: a^3\left[(1-\Omega_b^2)R_4(a)+\frac{3}{2}m_0^2
\Omega_b^2{\dot{b}}^2 -6V_{r}(b)\right] + S_M,
\label{4d-action-2}
\ee
where 
\be
R_4(a)=-6\frac{\ddot{a}}{a}-6\frac{{\dot{a}}^2}{a^2},
\ee
and $S_{M}=\int d^4xa^3L^-\Omega_b^4$ is the matter action on the
negative tension brane. Notice that $R_4(a)$ is a Ricci scalar derived
from the metric $g_{\mu\nu}$ given in equation (\ref{hubble}). {}From
equation (\ref{conformal}), it is, therefore, obvious that
$R_4(a)=R_4^{(+)}(a)$. The four dimensional effective action on the
positive tension brane can then be easily read out from
(\ref{4d-action-2})\cite{CGRT}, viz., 
\be
S_{eff}^{(+)} = \int d^4x \sqrt{-g^{(+)}}\left[\frac{R_4^{(+)}(a)}
{16\pi G^{(+)}}+\frac{3m_0}{4\kappa^2}\Omega_b^2{\dot{b}}^2-
{3\over \kappa^2m_0}V_{r}(b)\right] + S_M \ ,
\label{4d+}
\ee
where $16\pi G^{(+)}=\frac{2\kappa^2m_0}{1-\Omega_b^2}=
[M_{pl}^{(+)}]^{-2}$ corresponds to the Newton's constant on the
positive tension brane. This action governs evolution of the positive
tension brane. We, however, are interested primarily in our (negative
tension) brane. To study the cosmology in our universe, we need to
determine the effective four dimensional action on the negative
tension brane.  Let us parametrize the metric on the negative tension 
brane by\cite{CGRT}
\be
g^{(-)}_{\mu\nu} = [-1, Y^{2}(\tau), Y^{2}(\tau), Y^{2}(\tau)] \ ,
\label{metric-}
\ee
where $\tau$ is the new time coordinate on the negative tension brane. {}From
eq.(\ref{conformal}), it is easy to see that $\tau$ 
is related to the time $t$ on the positive tension brane through 
\be
d\tau = e^{-m_{0}b/2}dt,\qquad
Y = e^{-m_{0}b/2} a \ .
\label{cosmic}
\ee
Using these relations in (\ref{4d-action}), we get the effective action on the
negative tension brane to be
\be
S_{eff}^{(-)} = \frac{3}{\kappa^{2}m_{0}} \int d^{3}xd\tau Y^{3}(\tau)
\left[\frac{(1-\Omega_{b}^{2})(\tau)}{\Omega_{b}^{2}(\tau)}
\frac{Y^{'2}(\tau)}{Y^{2}(\tau)} + 
\frac{\Omega_{b}^{'2}(\tau)}{\Omega_{b}^{4}(\tau)} - 
2\frac{Y^{'}(\tau)}{Y(\tau)}\frac{\Omega_{b}^{'}(\tau)}
{\Omega_{b}^{3}(\tau)}-{V_{r}(b)\over{\Omega_{b}^{4}}}\right] + S_{M} \ ,
\label{4d-}
\ee
where $S_{M}=\int d^{3}xd\tau Y^{3}(\tau)L^{-}$ is the action on the negative
tension brane in the new coordinate system and prime denotes differentiation
with respect to the argument.

This action can be brought to the Brans-Dicke form. To see this, we define a 
new variable $\phi$ by 
\be
\Omega_{b}^{2} = \frac{1}{1+\alpha\phi} \ ,
\label{warp}
\ee
where $\alpha=m_{0}/16M_{5}^3\pi$. It follows, after a straightforward
calculation, that the action (\ref{4d-}) can be recast in the form
\be
S_{eff}^{(-)} = \int \frac{1}{16\pi} \sqrt{-g^{(-)}(\tau)} d^{3}xd\tau 
\left[\phi(\tau)R_{4}^{(-)}(\tau) - 
\frac{\omega(\phi(\tau))}{\phi(\tau)}\phi^{'2}(\tau)-V(\phi)\right] + 
S_{M}^{(-)} \ ,
\label{BD}
\ee
where 
\bea
\sqrt{-g^{(-)}(\tau)} &=& Y^3(\tau), \\
R_{4}^{(-)}(\tau) &=& -6\frac{Y^{''}(\tau)}{Y(\tau)}-
6\frac{Y^{'2}(\tau)}{Y^{2}(\tau)},\\
V(\phi) &=& V_{r}(b)(1+\alpha\phi)^{2},\\
\omega(\phi(\tau)) &=& -\frac{3}{2}\frac{\alpha\phi(\tau)}
{1+\alpha\phi(\tau)}  \ .
\label{BD-Ricci}
\eea

The action (\ref{4d-}), rewritten as in
(\ref{BD}), corresponds to a generalised BD theory \cite{BD}, where
the BD parameter $\omega$ is a function of the BD scalar field
$\phi$. A few comments are in order at this point. Recall that the
function $\Omega_b$ is instrumental in providing the resolution of the
large mass hierarchy between the electroweak scale and the four
dimensional Planck scale. This can be achieved only if $\Omega_b$
takes very small value (of $o(10^{-16})$). It is, therefore, clear
from (\ref{warp}) that typical values that $\alpha\phi$ should take,
to conform to the expected behaviour of $\Omega_b$, will be quite
large (of $o(10^{16})$). An immediate implication of this is that
$\omega(\phi)$ stays mostly in the vicinity of $-3/2$. This, though, seems
to be in direct conflict with 
astronomical observations (particularly the Solar system 
experiments~\cite{WND}) which require $|\omega| > 3000$.
We will, however, see in the next
section that our solution to the equations of motion are such that the
Brans-Dicke scalar stabilizes to its equilibrium value well within the
inflationary epoch\footnote{Some of these results are similar to those 
obtained by Chiba~\protect\cite{chiba} in the static metric limit.}.

\section{Inflation and Graceful exit}

The four dimensional averaged action for the RS field, as we saw in
the previous section, gives us the Brans-Dicke action with a specific
form for $\omega(\phi)$ given in (\ref{BD-Ricci}). In this section, we
will write down the equations of motion and obtain a solution to
these equations of motion.  As mentioned in the last section, the
solution we wish to seek is the one which, in the static limit, reduces
to the Randall-Sundrum solution. While preserving the solution to
the hierarchy problem on the visible brane, this still
leaves open the possibility of
interesting phenomena in early cosmology. Before we venture to
obtain the solution to the equations of motion, we would like to
point out that,  as in the Brans-Dicke theory, the effective Newton's
constant in this case equals the inverse of the Brans-Dicke
scalar field $\phi$. Therefore, using our definition of $\phi$ as in
(\ref{warp}), we can write
\be
\phi = \frac{1}{G^{(-)}} = \frac{1}{\alpha}\frac{(1-\Omega_{b}^{2})}
{\Omega_{b}^{2}}.
\label{BD-scalar}
\ee
Viewing gravity on the branes as a four-dimensional theory, the corresponding
`Planck masses' are then given by
\be
M_{pl}^{(-) 2} = \frac{1}{2\kappa^2m_0}\frac{(1-\Omega_{b}^{2})}
{\Omega_{b}^{2}} = e^{m_{0}b} M_{pl}^{(+)2}.
\label{planck}
\ee
The relation between $M_{pl}^{(\pm)}$
is the same as obtained by RS and serves to explain the hierarchy
between the energy scales. To have the right size for this ratio,
one needs have
$m_0b_0\sim 70$ where $b_0$ is the stabilised value of the radion
$b$. In other words, for the stabilised situation, $\Omega_{b_0}$ is a very
small number. 
We will ignore the matter terms on the negative tension brane.
However, we will keep the bulk matter fields in our analysis. This
assumption is justified because time evolution of the bulk metric is
expected to be much slower than that on the brane which is undergoing
exponential inflation due to warp factor. Therefore, the brane matter
is getting inflated away while the bulk matter, which we will take to 
be independent of brane spatial coordinates, is not.

The equations of motion, as obtained from the action in (II.16),  are
\cite{BD}
\bea
G_{\mu\nu} &=& \frac{T_{\mu\nu}}{\phi}
        +\frac{\omega}{\phi^2} (\phi_{,\mu} \phi_{,\nu}
-\frac{1}{2}g_{\mu\nu}\phi_{,\alpha}\phi^{,\alpha})+\frac{1}{\phi}
[\phi_{_{,}\mu_{;}\nu}
-g_{\mu\nu}\square{\phi}] - g_{\mu\nu} \frac{V(\phi)}{2 \phi},
\label{BD-EQN}\\
&&(2\omega+3)\square\phi = T + \left[ \phi \frac{d V (\phi)}{d \phi} 
                                   - 2 V(\phi) \right],
\label{waveq}
\eea
where $T$ is the trace of energy momentum tensor $T_{\mu\nu}$ of the matter
fields on the brane. Neglecting the matter fields on the brane, 
the independent set of equations of motion are
\be
\barr{rcl}
3H^2 & = & \dis {\omega(\phi)\over 2}{\phi^{'2}(\tau)\over \phi^2(\tau)} -3H
{\phi^{'}(\tau)\over\phi(\tau)}+{V(\phi)\over 2\phi}\nonumber\\
2H'+3H^2 & = & \dis -{\omega(\phi)\over
2} \: {\phi^{'2}(\tau)\over\phi^2(\tau)}-{\phi^{''}(\tau)\over \phi(\tau)}
-2H{\phi'(\tau)\over\phi(\tau)} +{V(\phi)\over 2\phi}\label{BD-FRW}\\
\phi^{''}(\tau)+3H\phi^{'}(\tau) & = & \dis - {1\over 2\omega(\phi)+3}
\left[\phi^{'2}(\tau){d\omega(\phi)\over d\phi}+\phi{dV(\phi)\over
d\phi} - 2V(\phi) \right]
\earr
   \label{eq_of_motion}
\ee
where $H=\frac{Y'(\tau)}{Y(\tau)}$. It can be easily seen that only two
of these three equations are independent. For example, 
if we choose to work with the
first two, then it can be shown that the wave equation for the
BD scalar field (\ref{waveq}) is satisfied identically.

Let us now perform the  transformation
\be
\tilde g_{\mu\nu}=\phi g_{\mu\nu},\quad \tilde Y^2 = \phi 
Y^2, \quad d\tilde\tau^2 = \phi d\tau^2,
\ee
followed by a field redefinition
\be
\phi \rightarrow \psi \; : \quad \left({d\psi\over d\phi}\right)^2 = 
{2\omega(\phi)+3\over 4\phi^2} \ .
\ee
In other words,
\be
\psi = \pm
{\sqrt{3}\over{2}}\ \ln\left({\sqrt{1+\alpha\phi}+1\over{\sqrt{1+
\alpha\phi}-1}}\right),
          \label{psi_defn}
\ee
and for the sake of convenience, we shall work with the positive 
branch. Also, from now on, prime denotes differentiation w.r.t. $\tilde\tau$. 

The above serve to simplify 
the equations of motion considerably rendering them identical 
to those for a scalar coupled minimally to gravity: 
\bea
3\tilde H^2(\tilde\tau) &=& \psi^{'2}(\tilde\tau)+
V(\psi),\label{friedman}\\ 2\tilde H^{'}(\tilde\tau)+ 3\tilde
H^2(\tilde\tau) &=& - \psi^{'2}(\tilde\tau)+ V(\psi),\label{robertson}\\
\psi^{''}(\tilde\tau) + 3\tilde H
(\tilde\tau)\psi^{'}(\tilde\tau)&=& -{1\over 2}{dV\over
d\psi}.\label{walker} \eea
The above set of equations involve three unknowns, 
viz., $\psi(\tilde \tau)$, $\tilde H(\tilde \tau)$ and $V(\psi)$.
However, only two of the equations are independent. The system is thus 
underdetermined\footnote{Of course, if the stabilising potential were
specified, so would $V(\psi)$ be. And then the system is 
no longer underdetermined.}, and consequently we are
forced to fix one of them by making a suitable Ansatz. 
A convenient way of parametrizing the same is to posit 
that 
\be
{d\psi\over d\tilde\tau} = K(\psi)\label{ansatz}
\ee
where $K(\psi)$ is an as yet undetermined function of $\psi$. $K(\psi)$
can also be an explicit function of $\tilde\tau$. However, if $\psi$
is a monotonic function of $\tilde\tau$, which is a sensible
assumption at least for the inflationary era, then explicit
$\tilde\tau$ dependence of $K(\psi)$ can be reexpressed in terms of
its dependence on $\psi$. As we will see below this is precisely what
we observe. Even before we choose a
specific form for $K(\psi)$, we may further simplify our 
equations of motion.
Subtracting (\ref{friedman}) from (\ref{robertson}) and using
(\ref{ansatz}) gives
\be
{d\tilde H\over d\psi} = - K(\psi) \ ,
\qquad
V(\psi) = 3 \tilde H^2(\psi) - K^2(\psi)
\label{linde}
\ee

We are now in a position to specify an Ansatz for $K(\psi)$. 
Let us start with the simplest of relations, namely a monomial form
\be
K(\psi) = \beta\psi^\gamma,
        \label{eq:monomial}
\ee
where $\beta$ and $\gamma$ are constants. This immediately leads
to 
\be
\tilde H (\psi) = \Bigg\{
\barr{lcl}
        - \beta \left[ 
                        \frac{\psi^{\gamma + 1}}{\gamma + 1}  + H_0 \right]
        & \qquad & \gamma \neq -1
        \\[1.5ex]
        - \beta \left[ \ln \psi + H_0 \right]
        & \qquad & \gamma = -1
\earr
        \label{eq:hub}
\ee
and
\be
\psi = \Bigg\{
\barr{lcl}
        \psi_0 e^{\beta(\tilde\tau-\tilde\tau_{0})}
        & \qquad & \gamma = 1
        \\[1.5ex]
        \left[\beta (1 - \gamma) (\tilde\tau-\tilde\tau_{0})
              + \psi_0^{1 - \gamma}
                        \right]^{1 / (1 - \gamma)}
        & \qquad & \gamma \neq 1
\earr
        \label{eq:psi(t)}
\ee
where $ H_{0}$  and $\psi_{0}$ are constants of integration.
It follows then that, for $\gamma \neq -1$, 
\be
V(\psi) = \beta^2 \left\{ 3 \left[ \frac{\psi^{\gamma + 1}}{\gamma + 1} 
                                   + H_0 \right]^2 
                         - \psi^{2 \gamma}
                  \right\}.
        \label{eq:pot}
\ee
A potential bounded from below is achieved only for 
$\gamma > 0$ and this is a condition that we need to impose. 
Such a potential has extrema at the origin as well as at
$\psi_m$, which is a solution of 

\be
    \frac{\psi_{m}^{\gamma + 1}}{\gamma + 1} - \frac{\gamma}{3} 
\psi_{m}^{\gamma - 1}
                + H_0 = 0.
\label{eq:min}
\ee
As already pointed out, a resolution of the hierarchy problem needs 
$\Omega_b\simeq o(10^{-16})$, or, in other words, $m_{0}b_{0}\simeq  70$
at the present epoch. Since $\psi_m \sim \sqrt{3}\Omega_b$, it immediately 
follows that we can neglect the first term in the above equation, and
%
\be
H_{0} \simeq {\gamma\over{3}}\psi_{m}^{\gamma-1}.
\label{eq:appH}
\ee
The magnitude and sign of the vacuum energy
at the minimum in question depends crucially on $H_0$ and 
$\gamma$. As observations support a small positive value for the 
cosmological constant, this may be used to 
eliminate large regions of values of $\gamma$.
In view of the recent interest in tachyon-driven inflation, let 
us reconsider $\gamma < 0$. However, this immediately leads to 
a very large and negative value for $H_0$ (as long as we want a 
small $\psi_m$) and is therefore severely disfavoured. For 
$\gamma =0$ (when the potential is quadratic and hence seems to resemble
the chaotic inflation scenario), we again get a negative, albeit small, 
value for $H_0$. 
For $0<\gamma<1$, on the other hand, we have a
reversal of the earlier situation. Now $H_0$ to be positive but very large 
positive which, again, is not a desireable feature. 
The point $\gamma =1$ is special as this is associated with
$H_0\sim 1/3$ which implies small positive vacuum energy.
And finally, for $\gamma > 1$ it is easy to see that the
extremum of the potential at a small $\psi_m$ is actually a
maximum and hence does not represent a real vacuum. 
Thus, phenomenological considerations restrict us to the case of 
$\gamma =1$, and we now examine it in greater detail.

Substituting $\gamma =1$ in eqs.(\ref{eq:hub}--\ref{eq:pot}), we have
\be
\barr{rcl}
\psi(\tilde \tau) & = & \dis \psi_{0}e^{\beta(\tilde \tau-\tilde\tau_{0})},
\\[2ex]
H(\psi) & = & \dis -\beta[\psi^{2}/2+H_{0}],\\[2ex]
V(\psi) & = & \beta^{2}[3H_{0}^2+(3H_{0}-1)\psi^2+{3\over{4}}\psi^4].
\earr
      \label{gamma_eq_1}
\ee
%
Also from (\ref{eq:appH}) one gets $H_{0}\approx 1/3$. 

While we may continue to work with the field $\psi$ and
eqs (\ref{friedman})-(\ref{walker}), the 
calculation of the different parameters characterizing inflation and 
the comparison with the observational data is better achieved if 
we recast the physics in terms of a (conventional) dimensionful 
field $\chi$. We define, 
\be
\chi = \sqrt{2} M \psi  \label{defn_for_chi}
\ee
where $M= M_{pl} / \sqrt{8\pi} \sim 0.2M_{pl}$ is the reduced 
Planck mass.  At the minimum, then, $\chi_{m} \sim 5.3\times 10^{2} \gev$.
In terms of $\chi$, we may rewrite the evolution equations as 
\be
\barr{rcl}
3H^{2} & = & \dis {1\over{M^{2}}}
                 \left[ {\chi^{\prime 2} \over 2} +V(\chi)\right], \\[2.5ex]
2H^{\prime}+3H^{2} & = & \dis {1\over{M^{2}}}
                 \left[ - {\chi^{\prime 2} \over 2} + V(\chi)\right], \\[2ex]
\chi^{\prime\prime}+3H\chi^{\prime}& = & \dis -{dV\over{d\chi}}.
     \label{eqns_in_chi}
\earr
\ee
In the above, the potential has been redefined to be 
\be
V(\chi) \equiv M^2 V(\psi) 
         \label{potential_in_chi} \ .
\ee
Suppressing, for the moment, the quadratic terms, the potential now 
reads 
\be
V(\chi) \approx 3M^{2}\beta^{2} 
          \left[ H_{0}^{2}+{1\over{16}}{\chi^{4}\over{M^{4}}} \right].
\label{approx_pot}
\ee
Thus, only one unknown parameter, viz. $\beta$, remains 
in the potential and this we will 
fix by comparing our results with observations.

Apart from the Hubble parameter, cosmological expansion is also 
characterized by the deceleration parameter $q$:
\[
q=-\left[ {H^{\prime}\over{H^{2}}}+1\right] \ .
\]
For an inflating universe, $q<0$ and inflation stops at $q=0$. 
One could also investigate inflation in terms of 
the slow roll paramters which are defined as 
\be
\epsilon = {M^{2}\over{2 \: V^2}} 
           \left( { {\rm d} V \over{\rm d}\chi }\right)^{2}, \qquad
\eta = {M^{2}\over{V}} \; { {\rm d}^2 V \over{\rm d}\chi^2 } \ .
     \label{defn:eps_eta}
\ee
For slow roll inflation to occur,  both $|\epsilon| \ll 1$ and $|\eta| \ll 1$
need to be satisfied. 
If the slow roll conditions were valid, eqns.(\ref{eqns_in_chi}) can
be approximated to read
\be
3H^{2} = {1\over{M^{2}}}V(\chi), \qquad
-3H\chi^{\prime} = {{\rm d} V \over{\rm d}\chi } \ .
               \label{eqns_for_slowroll}
\ee
Within the region of validity of the above equations, 
one then has
\be
q \approx \epsilon-1,\label{new16}
\ee
and hence $q \leq 0$ is nearly equivalent to $\epsilon \leq 1$. 
Henceforth we will investigate 
the evoution of the universe in terms of $\epsilon $ 
as it renders easier comparison with different observational parameters.

Now, with the potential of eqn.(\ref{approx_pot}), one has
\be
\epsilon \approx 8 \left[{M \chi^3 \over \chi^4 + 16 H_0^2 M^4}
                     \right]^2, 
\qquad
\eta \approx {12 M^2 \chi^2 \over \chi^4 + 16 H_0^2 M^4} \ .
\label{new17}
\ee
Clearly, slow roll is possible for both $\chi \ll M$ and $\chi \gg M$.
In the former case, the field $\chi$ has to increase 
from a value $<5.3\times 10^{2} \gev$ and $\beta$ needs to be positive. 
But from the expression of $H$ in eqn.(\ref{gamma_eq_1}), one concludes 
that, in such a case the universe is actually collapsing. 
In other words, the scalar field in this model 
can not increase from a small value and yet be consistent with observations.

\begin{figure}[!ht]
\vspace {-25ex}
\centerline{\epsfxsize=10cm\epsfysize=12.0cm \epsfbox{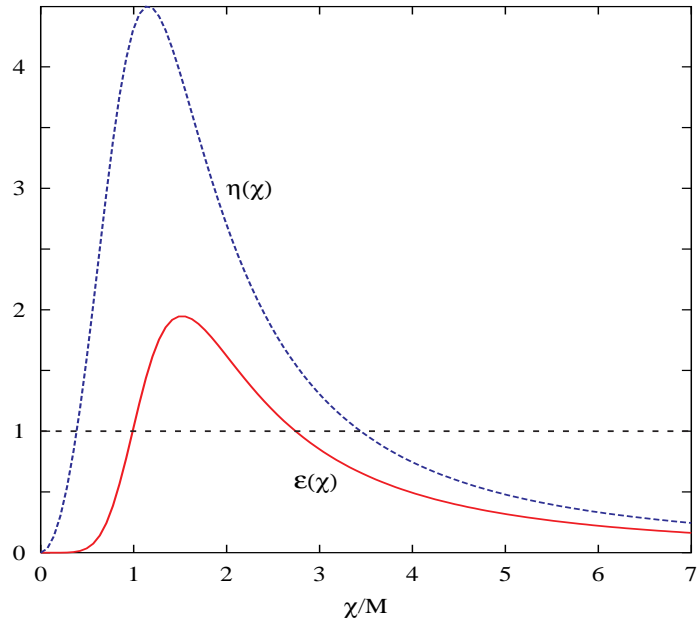}}
\caption{\sf The slow roll parameters $\eta$ (dashed line) and $\epsilon$
(solid line) as functions of the field $\chi$. The constant $H_0$ has 
been set to $1/3$.}
     \label{fig:decel}
\end{figure}
What if $\chi$ is actually decreasing from a large value?
For this to occur, $\beta < 0$ and eqn.(\ref{gamma_eq_1}) immediately 
implies that the universe is always expanding. In Fig.\ref{fig:decel},
we present the deceleration parameter as a function of the field $\chi$. 
For a large initial value $\chi_i$, the universe starts with an  
inflationary period which ends around $\chi_f \sim 3.44 M = 
0.69 M_{pl}$. 
The total amount of inflation is, of course, decided by the 
initial value $\chi_i$ and thus can be used to determine the latter.
This quantity is usually expressed as the number of e-foldings $N_e$ 
of inflation and can be quantified by
\be
N_e = - {1\over{M^{2}}}{\int^{\chi_{f}}_{\chi_{i}}} 
                d\chi \: V(\chi) \:
                \left({{\rm d} V \over{\rm d}\chi }\right)^{-1}
 \simeq {\chi_{i}^{2}-\chi_{f}^{2} \over{8 M^2}}
         \label{efoldings}
\ee
The minimal value of $N_e$, required for a satisfactory solution of 
the horizon and flatness problems, depends crucially on the energy 
scale of the inflationary expansion. Assuming that our inflation is 
a GUT scale one (an assumption we shall justify later), we 
must have had at least 60 e-foldings of inflation.
Using this minimal criterion (and the already determined value 
$\chi_{f} \sim 3.44 M $), one obtains $\chi_{i} \sim 4.42 M_{pl}$.
This superPlanckian starting point 
is quite reminiscent of the chaotic inflationary scenario.


Does our model lead to the right density perturbation spectrum? To answer 
this question, we need to consider two quantities, namely the 
spectral tilt $n_s$ 
\be
n_{s}=1-6\epsilon+2\eta\label{new20}
\ee
which describes the scale dependence of the perturbation and 
the ratio $r$ of the amplitude of tensor and scalar perturbations
\be
r \simeq 4 \pi A_{t}^{2}/A_{s}^{2} \ .
         \label{scalar_tensor_pert}
\ee
Both these quantities are to be evaluated at the 
instant when the comoving scale equals the Hubble 
radius ($k=aH$). For a GUT scale inflation with prompt reheating, 
this happens at around 55 e-foldings. Within our model, these 
quantities turn out to be
\be
	n_s \approx 0.95, \quad {\rm and} \quad 
	r \approx 0.12
\ee
and are very much consistent with the observational bounds 
inferred by the BOOMERANG \cite{BOOM}, MAXIMA \cite{maxima}, and 
DASI \cite{dasi} collaborations, namely
\be
0.8 < n_{s} < 1.05, \hspace{5mm} r < 0.3 \ .
\label{nsrsb}
\ee.
Note also that the spectral tilt is very similar 
to that for the Harrison-Zeldovich spetrum ($n_{s}=1$).

Before we end our discussion of the phenomenogical aspects of the model, 
we must comment on the as yet undetermined parameter $\beta$. It 
can be related to the amplitude of scalar perturbation:
\be
A_{s}^{2} = {512\pi V^{3} \over{75M_{pl}^{6}}} \: 
          \left({{\rm d} V \over{\rm d}\chi }\right)^{-2} \ .
\label{scalar_pert}
\ee
Observational data estimates $A_{s} \sim 2 \times 10^{-5}$ leading 
to a small value of $\beta \sim - 9.32 \times 10^{-8} M_{pl}$. Using 
this value of $\beta$ we have $V(\chi_f)^{1/4} \simeq 10^{15} \gev$, 
where $V(\chi_f)$ is the vacuum energy at the end of the inflation. 
This justifies our implicit assumption of GUT scale inflation.  

Could there be a second phase of inflation? A naive reading of 
Fig.~\ref{fig:decel} would cerainly seem to suggest that 
inflation could restart once $(\chi/M) \lsim 1$. 
But, in reality, this is not the case. 
Once the original phase of inflation ends, matter creation starts 
in right earnest. The presence of the matter terms in the 
r.h.s. of eqns.(\ref{eq_of_motion}) alters the dynamics radically 
thereby negating the possibility of a late inflation.

It is interesting to consider the dynamics of the field $b(\tilde \tau)$. 
Using eqns.(\ref{BD-scalar}), (\ref{psi_defn}) and (\ref{eq:psi(t)}), we have,
for $\gamma = 1$,
\be
m_0 b(\tilde \tau) = -2 
  \ln \tanh \left[ {2 \sqrt{\pi} \chi(\tilde \tau) \over \sqrt{3} M_{pl} }\right] \:
         = -2 \ln \tanh \left[{2 \sqrt{\pi} \chi_i e^{\beta \tilde \tau}
                          \over \sqrt{3} M_{pl} }\right]
           \label{eq:b}
\ee
where we denote the field at the initial instant of time by $\chi_i$. Note 
that while the first equality in eqn.(\ref{eq:b}) is but a definition (and 
hence valid for all times and all $\gamma$'s), the second equality 
is valid only when our approximation of neglecting the matter terms on 
the brane is a good one. While this approximation is an excellent one
during the inflationary era, it certainly is not so when matter creation
and reheating effects become important. 
Thus, although the ultimate settling down of the warp-factor to its 
current value is ensured, the time scale for the process cannot be 
determined from this analysis alone.
It is also interesting to note that 
the warp factor remains close to unity (and hence gravity on our brane 
is not suppressed in the least) during almost the entire inflationary 
era. Examining this issue quantitatively, 
although $b(t)$ changes by nearly 7 orders of magnitude 
during the inflationary era 
(from $m_{0}b(\chi_{i}) \simeq 5.5 \times 10^{-8}$ to   
$m_{0}b(\chi_{f}) \simeq 0.24$), this change is approximately only linear 
(vide eqn.\ref{eq:b}) in time\footnote{Subsequent evolution 
increases $b(t)$ by an additional factor of nearly 300
to reach $m_{0}b_{0} \simeq 70$.}. 
In other words, the expansion of the fifth dimension 
during inflation is far less than that of the observable brane. 
Consequently, the matter energy density on the 
brane, which, to begin with, was already much smaller than that in the 
bulk, is inflated away to a far greater degree than the latter. 
The asymmetric inflationary expansion of the 
higher dimensional world, thus, provides 
a further justification for ignoring the matter terms 
in the brane but not those in the bulk.
Furthermore, the relatively slow variation of the 
radion field $b(t)$ also explains the nearly scale-invariant nature of the 
primordial density perturbation which, in our model, 
arises primarily due to the fluctuations of the radion field. 

We now return to another important issue of our
scenario.  The result $|\omega|  \lsim 3/2$ seems to run counter to
the findings of the Solar system experiments~\cite{WND}.  However, it
needs to be realized that this model exits the Brans-Dicke phase at early 
times itself.  With $\psi$, and hence $\phi$, assuming a constant
value very quickly with time, it essentially decouples from the
equations of motion (see \ref{BD-EQN}).  And also as our BD field is 
massive, it no longer mediates
any long range force---at least classically--- and hence plays no role in
observables such as those considered in Ref.~\cite{WND} where one 
considered the massless BD field.\footnote{For other
interesting cosmological solutions to the Brans-Dicke gravity with this
behaviour of the scalar and the scale factor and with the late time
acceleration in the presence of a potential for the BD scalar, see
ref.\protect\cite{anjan_soma}.}. It might be
argued though that this line of reasoning depends crucially on our
having ignored the matter energy density and pressure.  Once the
variation of the BD scalar field $\phi$ becomes negligible with time
and particle creation processes take over, such terms may well become
dominant. Of course, the actual process of matter creation, which
takes place primarily through the processes of reheating and cooling,
depends on the details of the radion potential as well as the radion
couplings to the SM particles.  The latter, however, become relevant
only at very late times, and even then are not of a sufficiently large
magnitude to significantly alter the aforementioned results. 

\section{Summary}

To conclude, we exhibit that   a five-dimensional world with a  warped
geometry (a   time-dependent  analog  of the   Randall-Sundrum  metric
ansatz),   supports an inflationary evolution  on   the brane with the
desirable property of a graceful exit mechanism. The key ingredient is
the radion field whose coupling to the gravity on the negative tension
brane  is as that  of a  scalar field in  scalar tensor  gravity. By a
field transformation, we have recast  the action in a Brans-Dicke-like
form. The advantage of our approach over canonical models of inflation
is that  we do not need to  introduce any extra scalar (the inflaton). 
Rather, the radion itself drives  inflation.  Even more interestingly,
the radion evolution  exits   the inflationary phase in  a  completely
natural way, thus providing for a graceful exit.  We  have also made a
detailed analysis of the cosmological implications of our inflationary
model  by investigating  the density  perturabations  generated by the
radion. For this we have assumed a GUT scale  inflation which has also
been  justified by comparing the amplitude  of the scalar perturbation
in  our model  with the COBE  normalized  value. Two  concrete
predictions of our model are the values of the spectral index 
$n_{s}$ and the ratio of the tensor and scalar perturbations
$r$ which are both
well  within the range predicted by current  CMB observations. As a number
of new missions ({\sc map} and {\sc planck}) of 
CMB observations is under way, one
can  get a  more  constrained range for $n_{s}$ and $r$.  Since
$\gamma=1$  is  the only case  of  successful inflation in this model,
brane inflation based on RS type model  will be tested more accurately
by future observations.

\vspace*{2mm}
 
\noindent{\bf Acknowledgments}: We would like to thank Debashis Ghoshal,
Ashoke Sen and  Somasri Sen for  useful discussions.  We are extremely
thankful to the     referee  for detailed   report,    suggestions and
illuminating comments. SB thanks the Harish-Chandra Research Institute
for hospitality.  DC thanks the  Department of Science and Technology,
India  for financial assistance   under the  Swarnajayanti  Fellowship
grant.


%
\def\NPB#1#2#3{Nucl. Phys. {\bf B#1} #2 (#3)}
\def\PLB#1#2#3{Phys. Lett. {\bf B#1} #2 (#3)}
\def\PLBold#1#2#3{Phys. Lett. {\bf#1B} #2 (#3)}
\def\PRD#1#2#3{Phys. Rev. {\bf D#1} #2 (#3)}
\def\PRL#1#2#3{Phys. Rev. Lett. {\bf#1} #2 (#3)}
\def\PRT#1#2#3{Phys. Rep. {\bf#1} #2 (#3)}
\def\ARAA#1#2#3{Ann. Rev. Astron. Astrophys. {\bf#1} #2 (#3)}
\def\ARNP#1#2#3{Ann. Rev. Nucl. Part. Sci. {\bf#1} #2 (#3)}
\def\mpl#1#2#3{Mod. Phys. Lett. {\bf #1} #2 (#3)}
\def\ZPC#1#2#3{Zeit. f\"ur Physik {\bf C#1} #2 (#3)}
\def\APJ#1#2#3{Ap. J. {\bf #1} #2 (#3)}
\def\AP#1#2#3{{Ann. Phys. } {\bf #1} #2 (#3)}
\def\RMP#1#2#3{{Rev. Mod. Phys. } {\bf #1} #2 (#3)}
\def\CMP#1#2#3{{Comm. Math. Phys. } {\bf #1} #2 (#3)}
\relax
%
\newcommand{\journal}[4]{{ #1} {\bf #2}, #3 (#4)}
\newcommand{\hepth}[1]{{[hep-th/#1]}}
\newcommand{\hepph}[1]{{[hep-ph/#1]}}
\newcommand{\grqc}[1]{{[gr-qc/#1]}}
\newcommand{\astro}[1]{{[astro-ph/#1]}}
%


\end{document}